\documentclass[amsmath, twocolumn, showpacs, aps, prb]{revtex4-1}
\usepackage{bm}
\usepackage{graphics}
\usepackage{graphicx}
\usepackage{amssymb}
\usepackage{color}

\newcommand{\pdag}{{\phantom{\dagger}}}
\newcommand{\bsigma}{{\boldsymbol \sigma}}
\begin{document}

\title{Nuclear Spin Relaxation in Rashba Nanowires}
\author{Alexander A. Zyuzin, Tobias Meng, Viktoriia Kornich, and Daniel Loss}
\affiliation{
Department of Physics, University of Basel, Klingelbergstrasse 82, CH-4056 Basel, Switzerland
}


\begin{abstract}
We study 
%theoretically 
the nuclear spin relaxation  in a  ballistic nanowire
with hyperfine and Rashba spin-orbit  interactions (SOI) and in the presence of  magnetic field and  electron interactions. 
The relaxation rate shows pronounced peaks as function of magnetic field and chemical potential  due to van Hove singularities in the Rashba bands. As a result, 
the regimes of weak and strong SOIs can be distinguished by the
number of peaks in the rate. 
The relaxation rate increases with increasing magnetic field if both Rashba subbands are occupied, whereas it decreases  if  only the lowest one is occupied.

\end{abstract}
\maketitle

%{\emph{Introduction}}.~
\section{introduction}
Low-dimensional condensed matter systems with strong spin-orbit interaction (SOI) have attracted much attention both theoretically 
\cite{bib:Fu-Kane, bib:Sato, bib:Lutchyn, bib:Oreg, bib:Volovik-Review} and experimentally \cite{bib:ExpMF1, bib:ExpMF2, bib:ExpMF3, Rokhinson} 
for their realization of nontrivial momentum space topology. \cite{bib:Volovik-Review}
A particular example of such systems are semiconducting Rashba nanowires in a helical state, in which the Rashba SOI, \cite{Rashba} 
locks the spin of the electron to its direction of motion. In a Rashba nanowire, the helical state can be obtained by tuning the chemical potential into the partial gap at zero momentum induced by a magnetic field. Insulating and superconducting states of helical Rashba wires can host Jackiw-Rebbi \cite{bib:FF0, bib:Klinovaja_Stano} and Majorana bound states 
around topological defects. 
\cite{bib:AliceaReview}

Not only as a prerequisite for the  creation and identification of these exotic bound states, but also on its own right, it is important to gain
information about the strength of the SOI, and to detect signatures of the helical state in nanowires. For example, the drop of the conductance of a ballistic conduction channel from $2e^2/h$ to $e^2/h$ as a function of Fermi level can serve as an experimental probe of the helical state. \cite{StredaPRL2003,bib:goldhaber,Braunecker_2009,bib:HelicalExp} Signatures of the helical state can also be found in the electron spin susceptibility. \cite{meng_13_suscept} So far, the SOI in nanowires has been measured only in quantum dots (via transport) \cite{Fasth_2007,Kanai_2011,Nadj_2012}, where, however, the Rashba SOI  of interest is masked 
by the one that is induced by the dot confinement potential.

In this work, we propose an alternative and non-invasive way to access information about the SOI and the helical regime in a Rashba nanowire, namely via the nuclear spins. 
%and their relaxation. 
These are sensitive to the electronic state due to the hyperfine interaction present in III-V semiconductors such as GaAs or InAs.  

A main motivation for our proposal stems from striking experimental progress in the field of nanoscale magnetometry. In particular, it has recently been demonstrated that cantilever-based magnetic sensing enables the nuclear spin magnetometry of nanostructures, in particular of InP and GaP nanowires. \cite{bib:Poggio} In the remainder, we show how such ultra-sensitive techniques can be used to probe the strength of the SOI, and to detect the helical states  via the Korringa nuclear spin relaxation mechanism, i.e.~the change of the nuclear spin state due to the spin-flip scattering of itinerant electrons of energies within the thermally broadened region close to the Fermi level.  \cite{bib:Slichter} 
\begin{figure}[top]
  \centering
  \includegraphics[width=80mm]{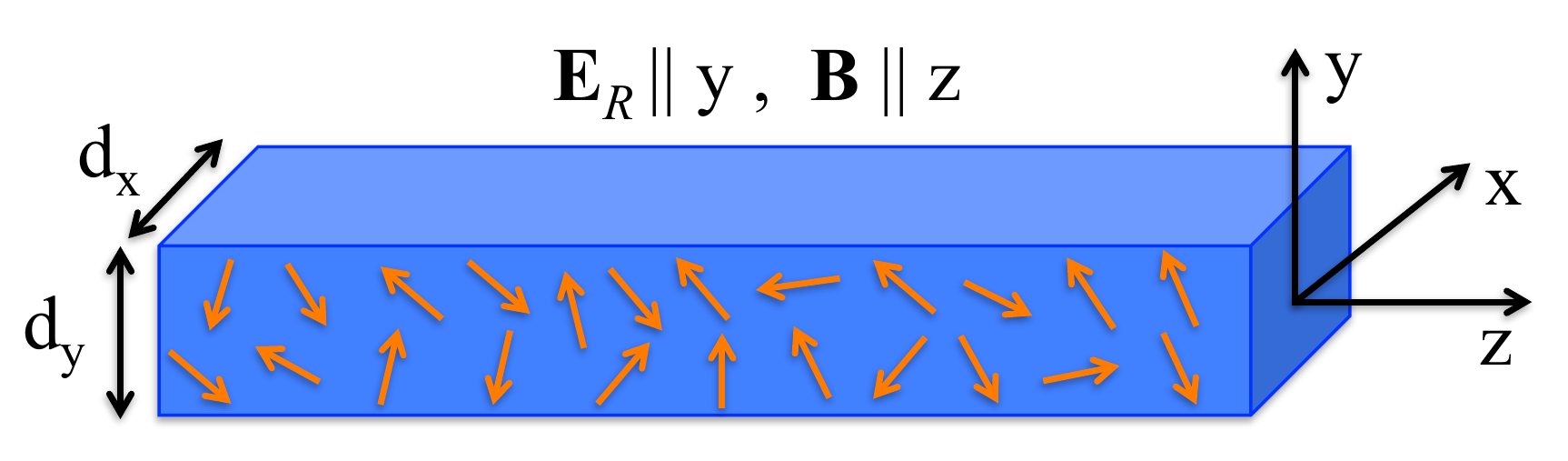}
\caption{Schematics of a nanowire of cross-sectional area, $d_x\times d_y$, confining a one-dimensional electron gas with Rashba electric field $\mathbf{ E}_R$ 
directed along $y$-axis and with magnetic field $\mathbf{ B}$ along $z$-axis. The itinerant electrons are coupled by hyperfine interaction to
localized nuclear spins (arrows).  
The Fermi wave-length $\lambda_F\approx d_x, d_y$ is much larger
than the  lattice spacing between nuclear spins.
}\label{fig1}
\end{figure}

We evaluate the nuclear spin relaxation rate in a one-dimensional ballistic electron gas in the presence of Rashba SOI and magnetic field.
We first derive an explicit dependence of the nuclear relaxation rate on the parameters of the electronic spectrum for non-interacting electrons, in which the relaxation rate is  proportional to the electronic temperature. We then discuss how electron interactions modify this temperature dependence to an interaction-dependent power law. 

We find that the relaxation rate shows distinct peaks as function of magnetic field and chemical potential  due to van Hove singularities in the Rashba bands.
Remarkably, 
the regime of weak SOI is characterized by one peak while the one of strong SOI by three peaks.
The  relaxation rate for weak SOI vanishes as a function of $\mu$ if  the Zeeman energy exceeds $\mu$. Strong SOI  gives rise to regions in the spectrum with negative group velocity and thus to pronounced peaks in the relaxation rate as function of $\mu$. 
Finally, we show that the relaxation rate increases with increasing Zeeman energy if the Fermi level crosses both Rashba bands, while the rate decreases  if only the lowest Rashba subband is occupied.

The outline of the paper is as follows. In Sec. II, we introduce the model Hamiltonian and derive the nuclear spin relaxation for non-interacting electrons for a Rashba band due to the hyperfine interaction. In Sec. III, we include electron-electron interactions described by a Luttinger liquid approach. In Sec. IV, we  give numerical estimates for the nuclear spin relaxation making use of the material parameters appropriate for In-based nanowires.

%{\emph{Relaxation in a non-interacting electron gas}}.~
\section{Relaxation in a non-interacting electron gas}
We consider nuclear spins coupled by hyperfine interaction to itinerant electrons in a semiconducting quantum or nano-wire with Rashba SOI. Our goal is to calculate 
the nuclear spin relaxation rate, first without electron-electron interactions.
We assume the electrons to occupy the lowest transverse subband of the wire with cross-sectional area $d_x\times d_y$, see Fig. (\ref{fig1}), described by the wave function $\Psi(x,y) = \frac{2}{\sqrt{d_x d_y}}\cos(\pi x/d_x)\cos(\pi y/d_y)$.
Taking into account that the nuclear density in the wire is much larger than the electron density we can
approximate the wave function $|\Psi(x,y)| \approx 2/\sqrt{d_xd_y}$ by its value at the centre of the wire and sum over the nuclear spin density in the transverse direction.
The Hamiltonian of the system becomes ($\hbar=1$)
\begin{widetext}
\begin{eqnarray}\label{Ham}
H=\int dz c^{\dag}_{s}(z)\left[-\left(\frac{\partial_z^2}{2m} +\mu\right)\delta_{ss'} -i\alpha\sigma_{ss'}^x\partial_z + h\sigma_{ss'}^z\right]c_{s'}(z)
+\int dz \bigg[\frac{\sqrt{d_xd_y}}{2}
A\mathbf{ I}(z)\cdot \frac{\bsigma_{ss'}}{2} c^{\dag}_{s}(z)c_{s'}(z)-\omega_N I^z(z)\bigg].
\end{eqnarray}
\end{widetext}
Here, $\mu$ is the chemical potential of the electrons with effective mass $m$ and spin projection $s$, where summation over repeated spin indexes is implied, $\alpha$ is the SOI constant, $h=g_e\mu_B B/2$ the Zeeman energy of the electrons due to the external magnetic field, $B$, applied along the $z$ axis, where $g_e$ is the electron g-factor and $\mu_B$ the Bohr magneton,  $\sigma^{x,y,z}$ the Pauli matrices, and 
$A=A_{3D} 4/(d_x d_y)$, where $A_{3D}$ is the bulk value of the hyperfine interaction constant between a nuclear spin and the spin of an electron.  We assume that the dominant contribution to the hyperfine
interaction is given by Fermi contact interaction. 

The nuclear spin density is given by $\mathbf{ I}(z) = N_{\perp}\sum_{j}\mathbf{ I}_{j}\delta(z-z_j)$, where $\mathbf{ I}_{j}$ is the spin operator of the $j$-th nuclei, $N_{\perp}$ is the number of nuclear spins in the  plane transverse to the wire axis, and the sum runs over the nuclear spins  the wire axis. We assume that the Zeeman energy of the nuclear spins induced by the external magnetic field, $\omega_N =g_N\mu_N B $, where $\mu_N$ and $g_N$ are the nuclear magneton and the effective g-factor, respectively, is small compared to the temperature $T$.
 We also assume temperature to be larger than the Kondo temperature associated with the localized spin.

It is convenient to express the single-particle Green function in the Rashba eigenbasis (for $A=0$),
\begin{eqnarray}
G_{ss'}(p, \epsilon_n) =\frac{1}{2}\sum_{\lambda=\pm 1} \frac{\delta_{ss'}-\lambda  \frac{\alpha p\sigma_{ss'}^x+h\sigma_{ss'}^z}
{\sqrt{\alpha^2 p^2+h^2}}}{i\epsilon_n-\epsilon_{\lambda}(p)},
\end{eqnarray}
where $\epsilon_{\lambda}(p) =p^2/2m -\mu -\lambda\sqrt{\alpha^2 p^2+h^2}$ is the electron spectrum and  $p$  the momentum
in the Rashba spin-split subband defined by $\lambda=\pm1$; $\epsilon_n=(2n+1)\pi T$ is the fermionic Matsubara frequency.

We can now calculate the nuclear spin relaxation rate $1/T_1$ in a one-dimensional electron gas with SOI and Zeeman energy 
in second order perturbation theory in the hyperfine interaction between the nuclear spin and electron spin density.
 The relaxation rate is determined by the dynamical spin susceptibility of the conduction electrons at the nuclear site as: \cite{bib:Slichter, bib:White}
\begin{eqnarray}
\frac{1}{T_{1}} =\lim_{\omega_{N}\rightarrow 0}\frac{2TA^2}{\omega_N}\int \frac{dq}{2\pi}\mathrm{Im}[\chi_{xx}(q,\omega_k)|_{i\omega_k\rightarrow \omega_N +i0^{+}}],~\label{eq:sucept_gen}
\end{eqnarray}
where $\omega_k =2\pi kT $ is the bosonic Matsubara frequency and we take the analytical continuation of the spin susceptibility, which for noninteracting electrons is given by,
\begin{eqnarray}\nonumber
\chi_{ab}(q,\omega_k)&=&-\frac{T}{4}\int \frac{dp}{2\pi}\sum_{n}G_{s s_1}(p+q,\epsilon_n+\omega_k)\sigma_{s_1s_2}^a\\
&\times&
G_{s_2 s_3}(p,\epsilon_n)\sigma_{s_3 s}^b\, .\label{eq:suscept}
\end{eqnarray}
Summing over
spin indexes and Matsubara frequencies and taking the limit  $\omega_N\rightarrow 0$ we obtain,
\begin{eqnarray}\label{T1General}
\frac{1}{T_{1}} &=& - \frac{\pi A^2T}{8}\sum_{\lambda,\lambda'} \int \frac{dpdq}{4\pi^2}\delta(\epsilon_{\lambda}(p)-\epsilon_{\lambda'}(q))\frac{\partial n_F(\epsilon)}{\partial \epsilon}\bigg|_{\epsilon=\epsilon_{\lambda}(p)}\\\nonumber
&\times&\bigg[\delta_{ss_1}-\frac{\lambda h\sigma^z_{ss_1}}{\sqrt{\alpha^2p^2+h^2}}\bigg]\sigma^x_{s_1s_2}\bigg[\delta_{s_2s_3}-\frac{\lambda' h\sigma^z_{s_2s_3}}{\sqrt{\alpha^2q^2+h^2}}\bigg]\sigma^x_{s_3s},
\end{eqnarray}
where $n_F(\epsilon)$ is the Fermi distribution function.
\begin{figure}[top]
  \centering
  \begin{tabular}{cc}
   \includegraphics[width=80mm]{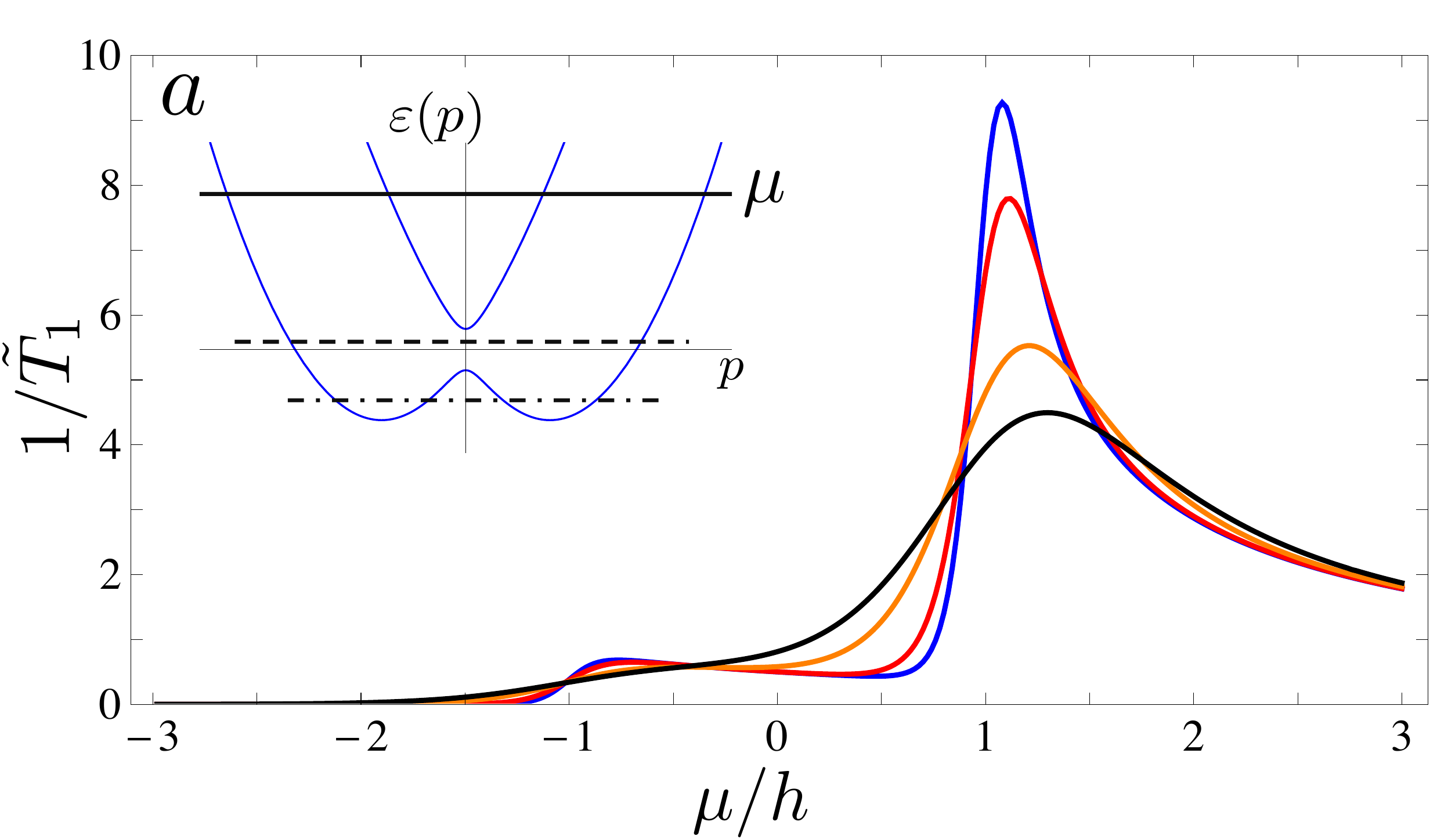}\\
       \includegraphics[width=80mm]{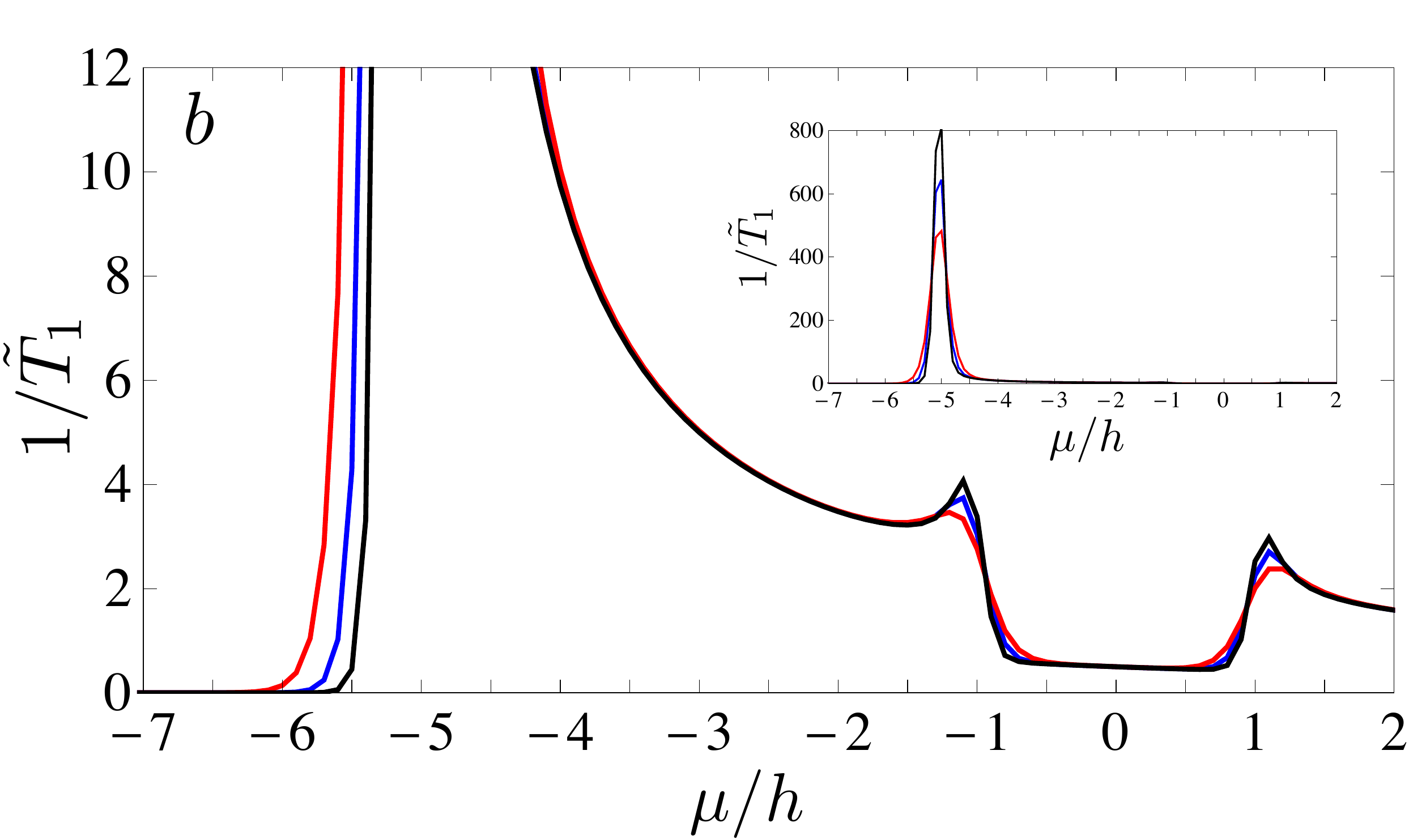}\\
  \end{tabular}\caption{The nuclear spin relaxation rate $1/\tilde{T}_1 = T_1(0)/(2T_1(\mu/h))$ 
as  function of  $\mu/h$ 
for small (a) and large (b) SOI strengths $m\lambda^2/h =0.2$ and $m\lambda^2/h =10$, respectively, as numerically obtained from Eq.~(\ref{T1General}).
  The  rate is normalized by its value at $\mu=0$ (we set $\mu=0$ at the
 middle of the Zeeman gap  between the  Rashba subbands at $p=0$).
The  pronounced peaks are due to van Hove singularities at the edges of the Rashba subbands.
(a) Curves from top to bottom are for $T/h = 0.07, 0.1, 0.2, 0.3$, respectively. (b) Curves from top to bottom are for $T/h = 0.05, 0.07, 0.1$, respectively. Inset (a): The Rashba spectrum  with possible Fermi levels (lines). Strong SOI gives rise to regions in the spectrum (dashed-dotted line) with negative group velocity and non-monotonic relaxation rate as shown in (b). 
The strong increase of the rate for $\mu/h$ approaching the band bottom $E_0/h=-5$ (see also inset (b)) signals the breakdown of perturbation theory.
}\label{fig2}
\end{figure}

\begin{figure}[top]
  \centering
  \begin{tabular}{cc}
   \includegraphics[width=80mm]{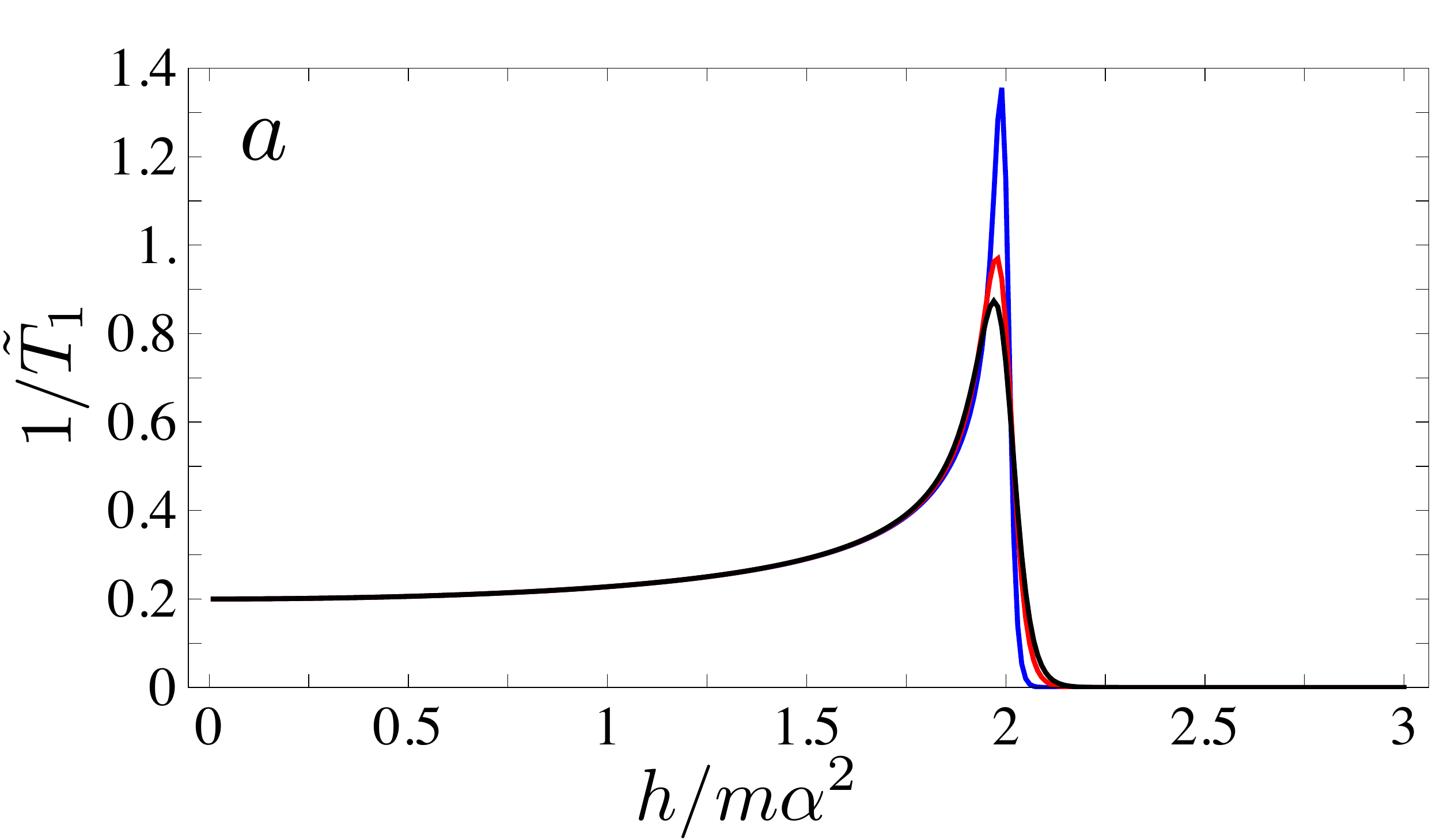}\\
       \includegraphics[width=80mm]{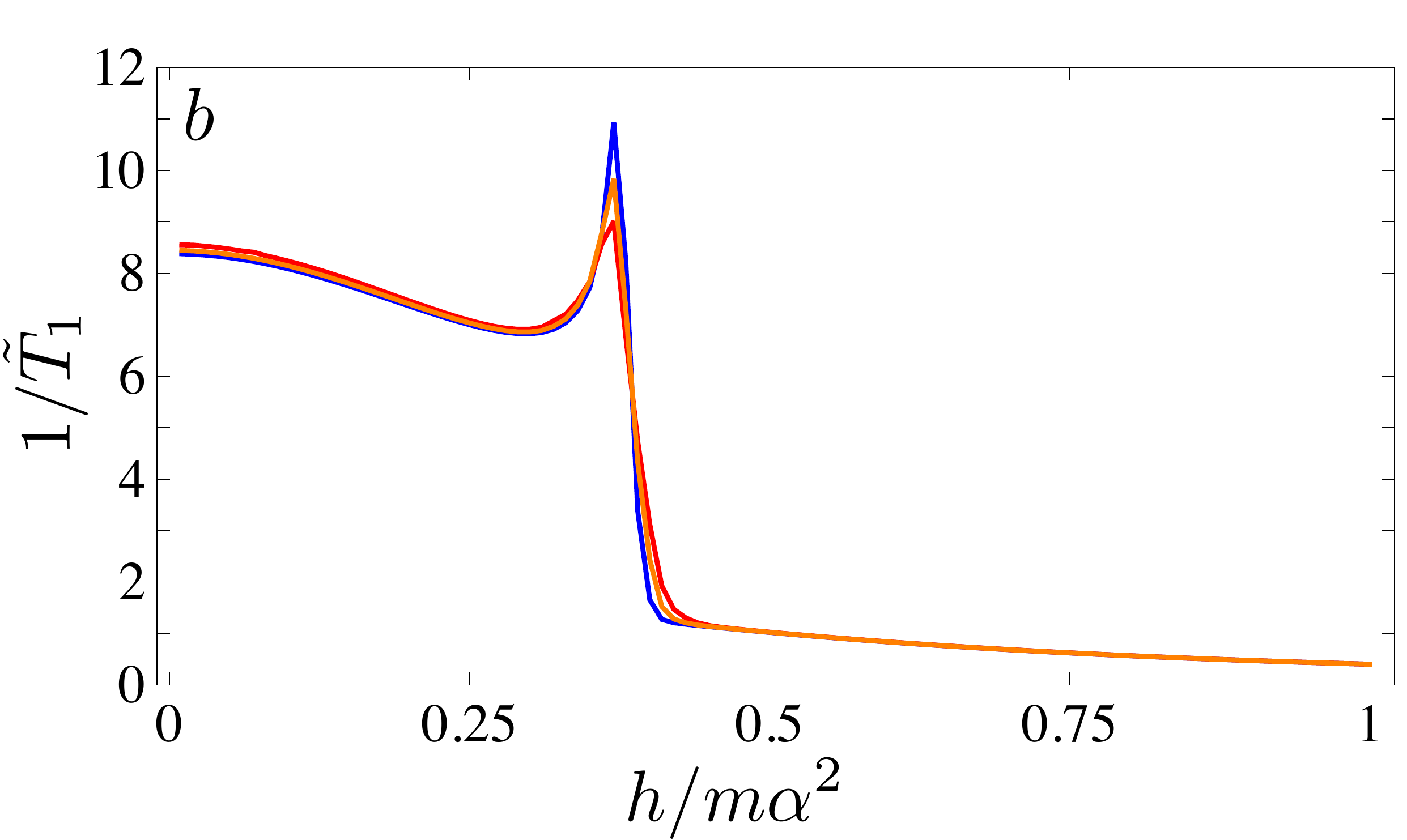}\\
  \end{tabular}\caption{Nuclear spin relaxation rate $1/\tilde{T}_1 = T_1(h=0)_{\mu=0}/(
  2T_1(h/m\alpha^2))$ as a function of Zeeman energy $h$ normalized by the SOI energy $h/m\alpha^2$ for two cases:
  (a) $\mu/m\alpha^2=2$,  $T/m\alpha^2 =0.01, 0.02, 0.025$. The corresponding Fermi level is shown by the solid line in Fig. \ref{fig2}(a).
  (b) $\mu/m\alpha^2= -0.38$,  $T/m\alpha^2 =0.005,0.0075,0.01$. The corresponding Fermi level is shown by the dashed-dotted line in Fig. \ref{fig2}(a). Note that the singularities are smoothened into finite peaks by temperature effects.
}\label{fig3}
\end{figure}

We integrate over momentum and obtain 
the relaxation rate in the linear temperature regime as,
\begin{eqnarray}\label{T1}
\frac{1}{T_{1}}&=&\frac{\Theta(h^2+m^2\alpha^4+2m\alpha^2\mu)}{h^2+m^2\alpha^4+2m\alpha^2\mu}
\bigg\{
m\alpha^2 (\Theta(E_+)+\Theta(E_-))\\\nonumber
&+&2\frac{\Theta(\mu^2-h^2)}{\sqrt{\mu^2-h^2}}(h^2+m\alpha^2 \mu)\mathrm{sign}(h^2+2m\alpha^2\mu)\bigg \} \frac{mA^2T}{4\pi},
\end{eqnarray}
where $E_{\pm} = \mu +m\alpha^2\pm\sqrt{h^2+m^2\alpha^4+2m\alpha^2\mu}$; $\Theta(x)$ and $\mathrm{sign}(x)$ are the Heaviside and sign functions, respectively. 
The dependencies of the relaxation rate on the chemical potential $\mu$ and the Zeeman energy $h$ are plotted in Figs. \ref{fig2} and \ref{fig3}, respectively.

Let us now consider the relaxation rate in several limiting regimes of the electronic spectrum. 
Without the SOI, {\it i.e.}, $\alpha=0$, we obtain,
\begin{equation}
\frac{1}{T_{1}} =\frac{mA^2T}{2\pi}\frac{\Theta(\mu^2-h^2)}{\sqrt{\mu^2-h^2}},
\label{eq:helical}
\end{equation}
which in the absence of a magnetic field, $h=0$, leads to the well-known result for the Korringa relaxation rate, 
$T^{-1}_{K} =\pi TA^2\nu^2$, where $\nu = 1/\pi v_F$ is the density of states per spin in the one-dimensional electron gas. If the chemical potential is smaller than the Zeeman energy, $|\mu|< |h|$, then the relaxation rate $1/T_1$ vanishes. Physically, this expresses the fact that the nuclear spin polarization cannot decay via flip-flop processes with the electrons if the latter are spin polarized due to the presence of a large Zeeman field. We note that the present calculation does not take into account competing nuclear spin relaxation mechanisms.

In the vicinity of the van-Hove singularity, $\mu\gtrsim h>0$, see Fig. \ref{fig2}, the relaxation rate scales with the chemical potential as 
$T^{-1}_{1} =\frac{mA^2T}{4\pi}\frac{\Theta(\mu-h)}{h+m\alpha^2}\sqrt{\frac{2h}{\mu-h}}$.

On the other hand, in the presence of both SOI and magnetic field, tuning the chemical potential 
to the middle of the gap of the spectrum at $p=0$, $\mu=0$, we obtain,
\begin{equation}
\frac{1}{T_{1}} =\frac{mA^2T}{4\pi}\frac{m\alpha^2}{h^2+m^2\alpha^4}~.
\label{eq:zeeman}
\end{equation}
We note that the relaxation rate diverges at the van Hove singularities of the spectrum occurring at zeroes of $\partial \epsilon_{\lambda}(p)/\partial p$. For instance,
$1/T_1 \sim [h(\mu - h)]^{-1/2}$  for weak SOI and $\mu>h$, and
$1/T_1 \sim |\mu-E_0|^{-1}$ for strong SOI, where $E_0$ denotes the band bottom.
Formally, the perturbation expansion in $A$ for the rate breaks down at these singularities. However,
these singularities  turn into well-defined peaks by 
finite-temperature effects, as we confirmed by evaluating Eq.~(\ref{T1General}) numerically for various temperatures, see Figs. \ref{fig2} and \ref{fig3}.
The peak at $E_0$, however, remains  large also for $T>0$ and thus is outside the perturbative regime considered here.
As an important result, we see that the relaxation rate behaves qualitatively very different for weak and for strong SOIs: in the former case, there is
only one peak, while in the latter there are three peaks in $1/T_1$ as function of $\mu$, see Figs. \ref{fig2}(a) and \ref{fig2}(b). 

Finally, let us discuss the dependence of the relaxation rate on the magnetic field $h$. The relaxation rate increases with the increase of the Zeeman energy as $\sim h^2/\mu^2$ for $\mu> m\alpha^2$ if the Fermi level crosses both Rashba subbands, see Fig. \ref{fig3}(a) and inset in Fig. \ref{fig2}(a), where the position of the Fermi level is shown by the solid line. 
On the other hand, the relaxation rate decreases with increase of the Zeeman energy as $\sim -h^2/m^2\alpha^4$ for $|\mu|< m\alpha^2$ if the Fermi level crosses only the lowest Rashba subband, see Fig. \ref{fig3}(b) and inset in Fig. \ref{fig2}(a), where the position of the Fermi level is shown by the dashed-dotted line. The singularity of the relaxation rate shown in Figs. \ref{fig3}(a) and \ref{fig3}(b) corresponds to the condition when $h=|\mu|$.

%{\emph{Relaxation in an interacting electron system}}.~ 
\section{Relaxation in an interacting electron system}
In a one-dimensional system, electron-electron interactions modify the temperature dependence of the nuclear spin relaxation rate from a linear scaling to interaction dependent power laws. We derive these in a Luttinger liquid calculation valid when the chemical potential is sufficiently far from the van Hove singularities, so that the dispersion can be linearized. For $|\mu| \gg |h|,\, m\alpha^2$, the system can be understood as a spinful Luttinger liquid with subleading corrections due to the SOI and the magnetic field. Neglecting these subleading terms, we first decompose the fermionic operators into their right and left moving parts, $c_s(z) \approx e^{iz  k_F} R_s(z)+e^{-iz  k_F} L_s(z)$, where $k_F$ is the Fermi momentum. Next, we bosonize these operators using standard techniques \cite{giamarchi_book} as $r_{s}(z) = U_{rs}/\sqrt{2\pi a_0}\,{\rm exp}(-i(r\phi_c(z) + rs \phi_\sigma(z)-\theta_c(z)- s \theta_{\sigma}(z))/\sqrt{2})$, where $U_{rs}$ is a Klein factor, and $a_0$ is the short distance cutoff of the Luttinger liquid theory (this cutoff is of the order of the lattice constant). The parameters $r=R,L\equiv +1,-1$ and $s=\uparrow,\downarrow\equiv +1,-1$ denote the direction of motion and the spin, respectively, while the bosonic fields $\phi_i$  are proportional to the integrated charge ($i=c$) or spin $(i=\sigma)$ density, $\theta_i$ is proportional to the integrated charge or spin current, and $[\phi_{i} (z), \theta_{j}(z') ]=\delta_{ij}(i\pi/2) {\rm sgn}(z'-z)$. The effect of electron-electron interactions is captured by a Luttinger liquid parameter $1\geq K_c\geq0$ for the charge fluctuations (we take the Luttinger parameter in the spin sector to be $K_\sigma\approx1$), such that the Luttinger liquid Hamiltonian reads $H = \int \frac{dz}{2\pi}\,\sum_{i=c,\sigma}\left[\frac{u_{i}}{K_{i}}(\partial_{z}\phi_{i})^2+u_{i}K_{i}(\partial_{z}\theta_{i})^2\right]$, where $u_c$ ($u_\sigma$) is the effective velocity in the charge (spin) sector. This Hamiltonian in turn allows us to calculate the relaxation rate $T_1^{-1}$ for interacting electrons. We first bosonize the real space, imaginary time susceptibility $\chi_{xx}(z,\tau)$ according to the above prescription, which yields

\begin{align}
\chi_{xx}(z,\tau) &= \sum_{r,r'}\frac{\langle T_\tau r_\uparrow^\dagger(z,\tau) r_\downarrow'(z,\tau) r'_\downarrow{}^\dagger(0,0) r_\uparrow^\pdag(0,0)\rangle}{4}+ \rm{h.c.}\nonumber\\
&=\sum_{r,r'}\frac{-e^{-i(r-r')z k_F }}{4(2\pi a_0)^2}\langle T_\tau\, e^{i(\tilde{\phi}_{rr'}(z,\tau)-\tilde{\phi}_{rr'}(0,0))}\rangle\nonumber\\ 
&+ \rm{h.c.}\, ,\label{eq:susept_boson}
\end{align}
with $\tilde{\phi}_{rr'}(z,\tau) =( (r-r')\phi_c(z,\tau)+(r+r')\phi_\sigma(z,\tau)-2\theta_\sigma(z,\tau))/\sqrt{2}$.
From this expression, the retarded spin susceptibility as a function of frequency can be calculated by virtue of a Wick rotation to real time, and a subsequent Fourier transformation.\cite{giamarchi_book} Eq.~\eqref{eq:sucept_gen} then yields the relaxation time as

\begin{align}
 \frac{1}{T_{1}}&\approx \mathcal{T}(\mu,K_c)\,T+\widetilde{\mathcal{T}}(\mu,K_c)\,T^{K_c}~.\label{eq:ll_gapless}
\end{align}
Here, $\mathcal{T}(\mu,K_c)$ and $\widetilde{\mathcal{T}}(\mu,K_c)$ are prefactors which in general depend on the short distance cutoff $a_0$, and which reduce to $\mathcal{T}(\mu,1)=\widetilde{\mathcal{T}}(\mu,1) = \frac{mA^2}{2\pi\mu}$ in the non-interacting limit. This is the expected scaling for the Korringa law in a Luttinger liquid, in which forward scattering gives rise to the term $\sim T$, while backscattering results in the contribution $\sim T^{K_c}$. \cite{giamarchi_book}

In the limit $m\alpha^2\gg |h| \gg |\mu|$, the system is to a good approximation helical (half of the spectrum is gapped, see Fig.~\ref{fig2}(a), while the gapless modes move in opposite directions and have antiparallel spins). In this case, all terms in Eq.~\eqref{eq:susept_boson} involving the gapped modes (i.e.~right and left movers of the ``wrong'' spin polarization) are negligibly small, while the contribution of the gapless helical modes can be derived from an effective spinless Luttinger liquid Hamiltonian. The latter can either be obtained by linearizing the gapless helical modes and then bosonizing these, or by starting from the full Luttinger liquid Hamiltonian including right and left movers of both spin species and integrating out the gapped degrees of freedom along the lines of Ref.~[\onlinecite{meng_13_suscept}]. In the latter case, the Luttinger liquid parameter of the helical modes relates to the ones in the spin and charge sectors as $K_{\rm hel} = 2 K_c\sqrt{u_c u_\sigma}/\sqrt{(u_c+K_cK_\sigma u_\sigma)(u_\sigma+u_cK_cK_\sigma)}$ (this implies $1\geq K_{\rm hel}\geq 0$), while their effective velocity is given by $\sqrt{u_c u_\sigma(u_c+u_\sigma K_cK_\sigma)}/\sqrt{u_\sigma+u_cK_cK_\sigma}$.\cite{meng_13_nucl} Using these effective parameters, we find that the contribution of the gapless modes to $\chi_{xx}(0,\tau)$ is: 

\begin{align}
\chi_{xx}(0,\tau) &= \frac{-1}{8(\pi a_0)^2}\frac{\left(\pi a_0 T/u_{\rm hel}\right)^{2K_{\rm hel}}}{[\sinh(i \pi T \tau)\sinh(-i \pi T \tau)]^{K_{\rm hel}}}~.
\end{align}
This implies that the relaxation rate takes the form, \cite{giamarchi_book}
\begin{align}
 \frac{1}{T_{1}}&\approx {\mathcal{T}}_{\rm hel}(\alpha,K_{\rm hel})\,T^{2K_{\rm hel}-1}~,\label{eq:ll_helical}
\end{align}
where $\mathcal{T}_{\rm hel}(\alpha,K_{\rm hel})$ 
is again a cutoff-dependent prefactor that reduces to $\mathcal{T}_{\rm hel}(\alpha,1)=\frac{A^2}{4\pi\alpha^2}$ in the non-interacting limit. This power law complies with the fact that in the helical regime the nuclear spin relaxation
results from electronic backscattering processes only. Quite remarkably, for strong interaction such that $K_{\rm hel} < 1/2$ the relaxation rate increases with decreasing temperature, in stark contrast to weak
or absent interactions where the rate decreases with decreasing temperature.

For $m\alpha^2\gg |\mu|\gg |h|$, the system is gapless, and thus shows a scaling of the type given in Eq.~\eqref{eq:ll_gapless}. For $|h|\gg |\mu|,\,m\alpha^2$, finally, the system essentially behaves like a spinful wire with a Zeeman splitting between spin up and spin down in which only the lower of the two Zeeman-split band is occupied. In this case, we obtain $T_{1}^{-1}\approx 0$,
which follows from Eq.~\eqref{eq:helical}, as well as from Eq.~\eqref{eq:zeeman} in the limit $m\alpha^2/h \to 0$. 

%{\emph{Conclusions}}.~
\section{Conclusions}
Let us now comment on the experimental observability of the predicted behavior of the nuclear spin relaxation rates in InAs nanowires.
For an $\textrm{InAs}$ nanowire with a cross-sectional area of $d_x\times d_y = 50\times 50$ $\textrm{nm}^2$,
and with Fermi velocity $v_F = 3\times 10^6$ $\textrm{cm}/\textrm{s}$, we obtain for the one-dimensional density of states 
$\nu = {1}/{\pi \hbar v_F} \approx 16$ $(\textrm{eV}\textrm{nm})^{-1}$.
For an electron $g$-factor $|g|=8$, the Zeeman energy $h\approx 4$ $ \mathrm{K}$ at a magnetic field of $1$ $\mathrm{Tesla}$, requires correspondingly low temperatures, $T\lesssim 4$ $\textrm{K}$. 
The dominant hyperfine coupling comes from In with nuclear spin $I=9/2$ and bulk constant $A_{3D}=3$ $\mu \textrm{eV} \textrm{nm}^3$.
With this, we estimate the nuclear spin relaxation time $T_K \approx (\frac{d_x d_y}{4\nu A_{3D} })^2\frac{\hbar}{\pi k_BT} \approx 380$ $\mathrm{s}$, at  $T=1\textrm{K}$. 
Remarkably, this estimate for $T_K$ is consistent with recent measurements performed on InP nanowires with cantilever techniques. \cite{bib:Poggio} 

%Competing phonon assisted relaxation mechanism \cite{bib:Abragam} can be ruled out at temperatures smaller than the Debye temperate, which
%is $280 \mathrm{K}$ in $\mathrm{InAs}$. 
The phonon-assisted relaxation mechanism can be distinguished by the temperature dependence of the relaxation rate, $1/T^{\mathrm{ph}}_1\sim T^{7}$ $(T^2)$ 
%and $1/T^{\mathrm{ph}}_1\sim T^2$ for temperatures smaller and larger than the Debye temperature, respectively \cite{bib:Abragam}, which is $280 \mathrm{K}$ in $\mathrm{InAs}$. 
for $T$ smaller (larger) than the Debye temperature,~\cite{bib:Abragam} which is $280 \mathrm{K}$ in $\mathrm{InAs}$. 
The effect of the nuclear dipole-dipole interaction on the nuclear relaxation
%nuclear spin polarization 
can be suppressed by small magnetic fields of the order of few $\mathrm{mT}$ when the nuclear Zeeman splitting is larger than the dipolar energy.~\cite{Maletinsky} 

Hence, we conclude that the hyperfine contact interaction is the most important term for describing nuclear spin relaxation in In-based nanowires with  $\mathrm{s}$-type conduction band at low temperatures.
%$\mathrm{InAs}$ nanowires. 
%As pointed out above, this conclusion is entirely consistent with recent measurements~\cite{bib:Poggio}, which report values for $T_K$
Again, this conclusion is  supported by recent experiments,~\cite{bib:Poggio} which measured values for $T_K$ of the same order as found above
for the hyperfine interaction.

The measurement of the Rashba and Dresselhaus SOI coefficients via weak antilocalization (WAL) effects in quasi-one dimensional  \textrm{InGaAs} wires was recently reported in Ref. \onlinecite{Sasaki}. These 750nm wide wires with several transverse conduction channels
were treated as quasi-one dimensional due to the fact that the spin-relaxation length is much larger than the width of the wire.
For a 1D wire with only a single conduction channel the WAL mechanism does not work, in contrast to the mechanism of the nuclear spin relaxation proposed here.
The nuclear relaxation rate measures directly the spectrum of the electrons via the density of states, while the WAL signal is more indirect, and could also be strongly affected
by `extrinsic' spin orbit effects, Elliot-Yafet effect, etc.

Thus, it seems worthwhile to search experimentally for the predicted signatures of the SOI in the relaxation rate as a function of magnetic field, chemical potential, and temperature.

%\emph{Acknowledgements}.~ 
\section{Acknowledgements}
We thank M. Poggio and D. Becker for discussions and acknowledge support from the Swiss NSF and NCCR QSIT.

\bibliography{Kor-Ref_PRB}

\begin{thebibliography}{29}%
\makeatletter
\providecommand \@ifxundefined [1]{%
 \@ifx{#1\undefined}
}%
\providecommand \@ifnum [1]{%
 \ifnum #1\expandafter \@firstoftwo
 \else \expandafter \@secondoftwo
 \fi
}%
\providecommand \@ifx [1]{%
 \ifx #1\expandafter \@firstoftwo
 \else \expandafter \@secondoftwo
 \fi
}%
\providecommand \natexlab [1]{#1}%
\providecommand \enquote  [1]{``#1''}%
\providecommand \bibnamefont  [1]{#1}%
\providecommand \bibfnamefont [1]{#1}%
\providecommand \citenamefont [1]{#1}%
\providecommand \href@noop [0]{\@secondoftwo}%
\providecommand \href [0]{\begingroup \@sanitize@url \@href}%
\providecommand \@href[1]{\@@startlink{#1}\@@href}%
\providecommand \@@href[1]{\endgroup#1\@@endlink}%
\providecommand \@sanitize@url [0]{\catcode `\\12\catcode `\$12\catcode
  `\&12\catcode `\#12\catcode `\^12\catcode `\_12\catcode `\%12\relax}%
\providecommand \@@startlink[1]{}%
\providecommand \@@endlink[0]{}%
\providecommand \url  [0]{\begingroup\@sanitize@url \@url }%
\providecommand \@url [1]{\endgroup\@href {#1}{\urlprefix }}%
\providecommand \urlprefix  [0]{URL }%
\providecommand \Eprint [0]{\href }%
\providecommand \doibase [0]{http://dx.doi.org/}%
\providecommand \selectlanguage [0]{\@gobble}%
\providecommand \bibinfo  [0]{\@secondoftwo}%
\providecommand \bibfield  [0]{\@secondoftwo}%
\providecommand \translation [1]{[#1]}%
\providecommand \BibitemOpen [0]{}%
\providecommand \bibitemStop [0]{}%
\providecommand \bibitemNoStop [0]{.\EOS\space}%
\providecommand \EOS [0]{\spacefactor3000\relax}%
\providecommand \BibitemShut  [1]{\csname bibitem#1\endcsname}%
\let\auto@bib@innerbib\@empty
%</preamble>
\bibitem [{\citenamefont {Fu}\ and\ \citenamefont {Kane}(2008)}]{bib:Fu-Kane}%
  \BibitemOpen
  \bibfield  {author} {\bibinfo {author} {\bibfnamefont {L.}~\bibnamefont
  {Fu}}\ and\ \bibinfo {author} {\bibfnamefont {C.~L.}\ \bibnamefont {Kane}},\
  }\href {\doibase 10.1103/PhysRevLett.100.096407} {\bibfield  {journal}
  {\bibinfo  {journal} {Phys. Rev. Lett.}\ }\textbf {\bibinfo {volume} {100}},\
  \bibinfo {pages} {096407} (\bibinfo {year} {2008})}\BibitemShut {NoStop}%
\bibitem [{\citenamefont {Sato}\ and\ \citenamefont
  {Fujimoto}(2009)}]{bib:Sato}%
  \BibitemOpen
  \bibfield  {author} {\bibinfo {author} {\bibfnamefont {M.}~\bibnamefont
  {Sato}}\ and\ \bibinfo {author} {\bibfnamefont {S.}~\bibnamefont
  {Fujimoto}},\ }\href {\doibase 10.1103/PhysRevB.79.094504} {\bibfield
  {journal} {\bibinfo  {journal} {Phys. Rev. B}\ }\textbf {\bibinfo {volume}
  {79}},\ \bibinfo {pages} {094504} (\bibinfo {year} {2009})}\BibitemShut
  {NoStop}%
\bibitem [{\citenamefont {Lutchyn}\ \emph {et~al.}(2010)\citenamefont
  {Lutchyn}, \citenamefont {Sau},\ and\ \citenamefont
  {Das~Sarma}}]{bib:Lutchyn}%
  \BibitemOpen
  \bibfield  {author} {\bibinfo {author} {\bibfnamefont {R.~M.}\ \bibnamefont
  {Lutchyn}}, \bibinfo {author} {\bibfnamefont {J.~D.}\ \bibnamefont {Sau}}, \
  and\ \bibinfo {author} {\bibfnamefont {S.}~\bibnamefont {Das~Sarma}},\ }\href
  {\doibase 10.1103/PhysRevLett.105.077001} {\bibfield  {journal} {\bibinfo
  {journal} {Phys. Rev. Lett.}\ }\textbf {\bibinfo {volume} {105}},\ \bibinfo
  {pages} {077001} (\bibinfo {year} {2010})}\BibitemShut {NoStop}%
\bibitem [{\citenamefont {Oreg}\ \emph {et~al.}(2010)\citenamefont {Oreg},
  \citenamefont {Refael},\ and\ \citenamefont {von Oppen}}]{bib:Oreg}%
  \BibitemOpen
  \bibfield  {author} {\bibinfo {author} {\bibfnamefont {Y.}~\bibnamefont
  {Oreg}}, \bibinfo {author} {\bibfnamefont {G.}~\bibnamefont {Refael}}, \ and\
  \bibinfo {author} {\bibfnamefont {F.}~\bibnamefont {von Oppen}},\ }\href
  {\doibase 10.1103/PhysRevLett.105.177002} {\bibfield  {journal} {\bibinfo
  {journal} {Phys. Rev. Lett.}\ }\textbf {\bibinfo {volume} {105}},\ \bibinfo
  {pages} {177002} (\bibinfo {year} {2010})}\BibitemShut {NoStop}%
\bibitem [{\citenamefont {Volovik}(2003)}]{bib:Volovik-Review}%
  \BibitemOpen
  \bibfield  {author} {\bibinfo {author} {\bibfnamefont {G.}~\bibnamefont
  {Volovik}},\ }\href@noop {} {\emph {\bibinfo {title} {The Universe in a
  Helium Droplet}}}\ (\bibinfo  {publisher} {Oxford University Press, Oxford},\
  \bibinfo {year} {2003})\BibitemShut {NoStop}%
\bibitem [{\citenamefont {Mourik}\ \emph {et~al.}(2012)\citenamefont {Mourik},
  \citenamefont {Zuo}, \citenamefont {Frolov}, \citenamefont {Plissard},
  \citenamefont {Bakkers},\ and\ \citenamefont {Kouwenhoven}}]{bib:ExpMF1}%
  \BibitemOpen
  \bibfield  {author} {\bibinfo {author} {\bibfnamefont {V.}~\bibnamefont
  {Mourik}}, \bibinfo {author} {\bibfnamefont {K.}~\bibnamefont {Zuo}},
  \bibinfo {author} {\bibfnamefont {S.~M.}\ \bibnamefont {Frolov}}, \bibinfo
  {author} {\bibfnamefont {S.~R.}\ \bibnamefont {Plissard}}, \bibinfo {author}
  {\bibfnamefont {E.~P. A.~M.}\ \bibnamefont {Bakkers}}, \ and\ \bibinfo
  {author} {\bibfnamefont {L.~P.}\ \bibnamefont {Kouwenhoven}},\ }\href@noop {}
  {\bibfield  {journal} {\bibinfo  {journal} {Science}\ }\textbf {\bibinfo
  {volume} {336}},\ \bibinfo {pages} {1003} (\bibinfo {year}
  {2012})}\BibitemShut {NoStop}%
\bibitem [{\citenamefont {Deng}\ \emph {et~al.}(2012)\citenamefont {Deng},
  \citenamefont {Yu}, \citenamefont {Huang}, \citenamefont {Larsson},
  \citenamefont {Caro},\ and\ \citenamefont {Xu}}]{bib:ExpMF2}%
  \BibitemOpen
  \bibfield  {author} {\bibinfo {author} {\bibfnamefont {M.~T.}\ \bibnamefont
  {Deng}}, \bibinfo {author} {\bibfnamefont {C.~L.}\ \bibnamefont {Yu}},
  \bibinfo {author} {\bibfnamefont {G.~Y.}\ \bibnamefont {Huang}}, \bibinfo
  {author} {\bibfnamefont {M.}~\bibnamefont {Larsson}}, \bibinfo {author}
  {\bibfnamefont {P.}~\bibnamefont {Caro}}, \ and\ \bibinfo {author}
  {\bibfnamefont {H.~Q.}\ \bibnamefont {Xu}},\ }\href@noop {} {\bibfield
  {journal} {\bibinfo  {journal} {Nano Lett.}\ }\textbf {\bibinfo {volume}
  {12}},\ \bibinfo {pages} {6414} (\bibinfo {year} {2012})}\BibitemShut
  {NoStop}%
\bibitem [{\citenamefont {Das}\ \emph {et~al.}(2012)\citenamefont {Das},
  \citenamefont {Ronen}, \citenamefont {Most}, \citenamefont {Oreg},
  \citenamefont {Heiblum},\ and\ \citenamefont {Shtrikman}}]{bib:ExpMF3}%
  \BibitemOpen
  \bibfield  {author} {\bibinfo {author} {\bibfnamefont {A.}~\bibnamefont
  {Das}}, \bibinfo {author} {\bibfnamefont {Y.}~\bibnamefont {Ronen}}, \bibinfo
  {author} {\bibfnamefont {Y.}~\bibnamefont {Most}}, \bibinfo {author}
  {\bibfnamefont {Y.}~\bibnamefont {Oreg}}, \bibinfo {author} {\bibfnamefont
  {M.}~\bibnamefont {Heiblum}}, \ and\ \bibinfo {author} {\bibfnamefont
  {H.}~\bibnamefont {Shtrikman}},\ }\href@noop {} {\bibfield  {journal}
  {\bibinfo  {journal} {Nat. Phys.}\ }\textbf {\bibinfo {volume} {8}},\
  \bibinfo {pages} {887} (\bibinfo {year} {2012})}\BibitemShut {NoStop}%
\bibitem [{\citenamefont {Rokhinson}\ \emph {et~al.}(2012)\citenamefont
  {Rokhinson}, \citenamefont {Liu},\ and\ \citenamefont {Furdyna}}]{Rokhinson}%
  \BibitemOpen
  \bibfield  {author} {\bibinfo {author} {\bibfnamefont {L.~P.}\ \bibnamefont
  {Rokhinson}}, \bibinfo {author} {\bibfnamefont {X.}~\bibnamefont {Liu}}, \
  and\ \bibinfo {author} {\bibfnamefont {J.~K.}\ \bibnamefont {Furdyna}},\
  }\href@noop {} {\bibfield  {journal} {\bibinfo  {journal} {Nature Physics}\
  }\textbf {\bibinfo {volume} {8}},\ \bibinfo {pages} {795} (\bibinfo {year}
  {2012})}\BibitemShut {NoStop}%
\bibitem [{\citenamefont {Rashba}(1960)}]{Rashba}%
  \BibitemOpen
  \bibfield  {author} {\bibinfo {author} {\bibfnamefont {E.~I.}\ \bibnamefont
  {Rashba}},\ }\href@noop {} {\bibfield  {journal} {\bibinfo  {journal} {Sov.
  Phys. Solid State}\ }\textbf {\bibinfo {volume} {2}},\ \bibinfo {pages}
  {1109} (\bibinfo {year} {1960})}\BibitemShut {NoStop}%
\bibitem [{\citenamefont {Jackiw}\ and\ \citenamefont {Rebbi}(1976)}]{bib:FF0}%
  \BibitemOpen
  \bibfield  {author} {\bibinfo {author} {\bibfnamefont {R.}~\bibnamefont
  {Jackiw}}\ and\ \bibinfo {author} {\bibfnamefont {C.}~\bibnamefont {Rebbi}},\
  }\href@noop {} {\bibfield  {journal} {\bibinfo  {journal} {Phys. Rev. D}\
  }\textbf {\bibinfo {volume} {13}},\ \bibinfo {pages} {3398} (\bibinfo {year}
  {1976})}\BibitemShut {NoStop}%
\bibitem [{\citenamefont {Klinovaja}\ \emph {et~al.}(2012)\citenamefont
  {Klinovaja}, \citenamefont {Stano},\ and\ \citenamefont
  {Loss}}]{bib:Klinovaja_Stano}%
  \BibitemOpen
  \bibfield  {author} {\bibinfo {author} {\bibfnamefont {J.}~\bibnamefont
  {Klinovaja}}, \bibinfo {author} {\bibfnamefont {P.}~\bibnamefont {Stano}}, \
  and\ \bibinfo {author} {\bibfnamefont {D.}~\bibnamefont {Loss}},\ }\href@noop
  {} {\bibfield  {journal} {\bibinfo  {journal} {Phys. Rev. Lett.}\ }\textbf
  {\bibinfo {volume} {109}},\ \bibinfo {pages} {236801} (\bibinfo {year}
  {2012})}\BibitemShut {NoStop}%
\bibitem [{\citenamefont {Alicea}(2012)}]{bib:AliceaReview}%
  \BibitemOpen
  \bibfield  {author} {\bibinfo {author} {\bibfnamefont {J.}~\bibnamefont
  {Alicea}},\ }\href@noop {} {\bibfield  {journal} {\bibinfo  {journal}
  {Reports on Progress in Physics}\ }\textbf {\bibinfo {volume} {75}},\
  \bibinfo {pages} {076501} (\bibinfo {year} {2012})}\BibitemShut {NoStop}%
\bibitem [{\citenamefont {Streda}\ and\ \citenamefont
  {Seba}(2003)}]{StredaPRL2003}%
  \BibitemOpen
  \bibfield  {author} {\bibinfo {author} {\bibfnamefont {P.}~\bibnamefont
  {Streda}}\ and\ \bibinfo {author} {\bibfnamefont {P.}~\bibnamefont {Seba}},\
  }\href@noop {} {\bibfield  {journal} {\bibinfo  {journal} {Phys. Rev. Lett.}\
  }\textbf {\bibinfo {volume} {90}},\ \bibinfo {pages} {256601} (\bibinfo
  {year} {2003})}\BibitemShut {NoStop}%
\bibitem [{\citenamefont {Quay}\ \emph {et~al.}(2010)\citenamefont {Quay},
  \citenamefont {Hughes}, \citenamefont {Sulpizio}, \citenamefont {Pfeiffer},
  \citenamefont {Baldwin}, \citenamefont {West}, \citenamefont
  {Goldhaber-Gordon},\ and\ \citenamefont {de~Picciotto}}]{bib:goldhaber}%
  \BibitemOpen
  \bibfield  {author} {\bibinfo {author} {\bibfnamefont {C.~H.~L.}\
  \bibnamefont {Quay}}, \bibinfo {author} {\bibfnamefont {T.~L.}\ \bibnamefont
  {Hughes}}, \bibinfo {author} {\bibfnamefont {J.~A.}\ \bibnamefont
  {Sulpizio}}, \bibinfo {author} {\bibfnamefont {L.~N.}\ \bibnamefont
  {Pfeiffer}}, \bibinfo {author} {\bibfnamefont {K.~W.}\ \bibnamefont
  {Baldwin}}, \bibinfo {author} {\bibfnamefont {K.~W.}\ \bibnamefont {West}},
  \bibinfo {author} {\bibfnamefont {D.}~\bibnamefont {Goldhaber-Gordon}}, \
  and\ \bibinfo {author} {\bibfnamefont {R.}~\bibnamefont {de~Picciotto}},\
  }\href@noop {} {\bibfield  {journal} {\bibinfo  {journal} {Nature Physics}\
  }\textbf {\bibinfo {volume} {6}},\ \bibinfo {pages} {336} (\bibinfo {year}
  {2010})}\BibitemShut {NoStop}%
\bibitem [{\citenamefont {Braunecker}\ \emph {et~al.}(2009)\citenamefont
  {Braunecker}, \citenamefont {Simon},\ and\ \citenamefont
  {Loss}}]{Braunecker_2009}%
  \BibitemOpen
  \bibfield  {author} {\bibinfo {author} {\bibfnamefont {B.}~\bibnamefont
  {Braunecker}}, \bibinfo {author} {\bibfnamefont {P.}~\bibnamefont {Simon}}, \
  and\ \bibinfo {author} {\bibfnamefont {D.}~\bibnamefont {Loss}},\ }\href@noop
  {} {\bibfield  {journal} {\bibinfo  {journal} {Phys. Rev. Lett.}\ }\textbf
  {\bibinfo {volume} {102}},\ \bibinfo {pages} {116403} (\bibinfo {year}
  {2009})}\BibitemShut {NoStop}%
\bibitem [{\citenamefont {Scheller}\ \emph {et~al.}(2014)\citenamefont
  {Scheller}, \citenamefont {Liu}, \citenamefont {Barak}, \citenamefont
  {Yacoby}, \citenamefont {Pfeiffer}, \citenamefont {West},\ and\ \citenamefont
  {Zumbuhl}}]{bib:HelicalExp}%
  \BibitemOpen
  \bibfield  {author} {\bibinfo {author} {\bibfnamefont {C.~P.}\ \bibnamefont
  {Scheller}}, \bibinfo {author} {\bibfnamefont {T.~M.}\ \bibnamefont {Liu}},
  \bibinfo {author} {\bibfnamefont {G.}~\bibnamefont {Barak}}, \bibinfo
  {author} {\bibfnamefont {A.}~\bibnamefont {Yacoby}}, \bibinfo {author}
  {\bibfnamefont {L.~N.}\ \bibnamefont {Pfeiffer}}, \bibinfo {author}
  {\bibfnamefont {K.~W.}\ \bibnamefont {West}}, \ and\ \bibinfo {author}
  {\bibfnamefont {D.~M.}\ \bibnamefont {Zumbuhl}},\ }\href@noop {} {\bibfield
  {journal} {\bibinfo  {journal} {Phys. Rev. Lett.}\ }\textbf {\bibinfo
  {volume} {112}},\ \bibinfo {pages} {066801} (\bibinfo {year}
  {2014})}\BibitemShut {NoStop}%
\bibitem [{\citenamefont {Meng}\ and\ \citenamefont
  {Loss}(2013)}]{meng_13_suscept}%
  \BibitemOpen
  \bibfield  {author} {\bibinfo {author} {\bibfnamefont {T.}~\bibnamefont
  {Meng}}\ and\ \bibinfo {author} {\bibfnamefont {D.}~\bibnamefont {Loss}},\
  }\href {\doibase 10.1103/PhysRevB.88.035437} {\bibfield  {journal} {\bibinfo
  {journal} {Phys. Rev. B}\ }\textbf {\bibinfo {volume} {88}},\ \bibinfo
  {pages} {035437} (\bibinfo {year} {2013})}\BibitemShut {NoStop}%
\bibitem [{\citenamefont {Fasth}\ \emph {et~al.}(2007)\citenamefont {Fasth},
  \citenamefont {Fuhrer}, \citenamefont {Samuelson}, \citenamefont {Golovach},\
  and\ \citenamefont {Loss}}]{Fasth_2007}%
  \BibitemOpen
  \bibfield  {author} {\bibinfo {author} {\bibfnamefont {C.}~\bibnamefont
  {Fasth}}, \bibinfo {author} {\bibfnamefont {A.}~\bibnamefont {Fuhrer}},
  \bibinfo {author} {\bibfnamefont {L.}~\bibnamefont {Samuelson}}, \bibinfo
  {author} {\bibfnamefont {V.~N.}\ \bibnamefont {Golovach}}, \ and\ \bibinfo
  {author} {\bibfnamefont {D.}~\bibnamefont {Loss}},\ }\href@noop {} {\bibfield
   {journal} {\bibinfo  {journal} {Phys. Rev. Lett.}\ }\textbf {\bibinfo
  {volume} {98}},\ \bibinfo {pages} {266801} (\bibinfo {year}
  {2007})}\BibitemShut {NoStop}%
\bibitem [{\citenamefont {Kanai}\ \emph {et~al.}(2011)\citenamefont {Kanai},
  \citenamefont {Deacon}, \citenamefont {Takahashi}, \citenamefont {Oiwa},
  \citenamefont {Yoshida}, \citenamefont {Shibata}, \citenamefont {Hirakawa},
  \citenamefont {Tokura},\ and\ \citenamefont {Tarucha}}]{Kanai_2011}%
  \BibitemOpen
  \bibfield  {author} {\bibinfo {author} {\bibfnamefont {Y.}~\bibnamefont
  {Kanai}}, \bibinfo {author} {\bibfnamefont {R.~S.}\ \bibnamefont {Deacon}},
  \bibinfo {author} {\bibfnamefont {S.}~\bibnamefont {Takahashi}}, \bibinfo
  {author} {\bibfnamefont {A.}~\bibnamefont {Oiwa}}, \bibinfo {author}
  {\bibfnamefont {K.}~\bibnamefont {Yoshida}}, \bibinfo {author} {\bibfnamefont
  {K.}~\bibnamefont {Shibata}}, \bibinfo {author} {\bibfnamefont
  {K.}~\bibnamefont {Hirakawa}}, \bibinfo {author} {\bibfnamefont
  {Y.}~\bibnamefont {Tokura}}, \ and\ \bibinfo {author} {\bibfnamefont
  {S.}~\bibnamefont {Tarucha}},\ }\href@noop {} {\bibfield  {journal} {\bibinfo
   {journal} {Nat. Nano}\ }\textbf {\bibinfo {volume} {6}},\ \bibinfo {pages}
  {511} (\bibinfo {year} {2011})}\BibitemShut {NoStop}%
\bibitem [{\citenamefont {Nadj-Perge}\ \emph {et~al.}(2012)\citenamefont
  {Nadj-Perge}, \citenamefont {Pribiag}, \citenamefont {van~den Berg},
  \citenamefont {Zuo}, \citenamefont {Plissard}, \citenamefont {Bakkers},
  \citenamefont {Frolov},\ and\ \citenamefont {Kouwenhoven}}]{Nadj_2012}%
  \BibitemOpen
  \bibfield  {author} {\bibinfo {author} {\bibfnamefont {S.}~\bibnamefont
  {Nadj-Perge}}, \bibinfo {author} {\bibfnamefont {V.~S.}\ \bibnamefont
  {Pribiag}}, \bibinfo {author} {\bibfnamefont {J.~W.~G.}\ \bibnamefont
  {van~den Berg}}, \bibinfo {author} {\bibfnamefont {K.}~\bibnamefont {Zuo}},
  \bibinfo {author} {\bibfnamefont {S.~R.}\ \bibnamefont {Plissard}}, \bibinfo
  {author} {\bibfnamefont {E.~P. A.~M.}\ \bibnamefont {Bakkers}}, \bibinfo
  {author} {\bibfnamefont {S.~M.}\ \bibnamefont {Frolov}}, \ and\ \bibinfo
  {author} {\bibfnamefont {L.~P.}\ \bibnamefont {Kouwenhoven}},\ }\href@noop {}
  {\bibfield  {journal} {\bibinfo  {journal} {Phys. Rev. Lett.}\ }\textbf
  {\bibinfo {volume} {108}},\ \bibinfo {pages} {166801} (\bibinfo {year}
  {2012})}\BibitemShut {NoStop}%
\bibitem [{\citenamefont {Peddibhotla}\ \emph {et~al.}(2013)\citenamefont
  {Peddibhotla}, \citenamefont {Xue}, \citenamefont {Hauge}, \citenamefont
  {Assali}, \citenamefont {Bakkers},\ and\ \citenamefont
  {Poggio}}]{bib:Poggio}%
  \BibitemOpen
  \bibfield  {author} {\bibinfo {author} {\bibfnamefont {P.}~\bibnamefont
  {Peddibhotla}}, \bibinfo {author} {\bibfnamefont {F.}~\bibnamefont {Xue}},
  \bibinfo {author} {\bibfnamefont {H.~I.~T.}\ \bibnamefont {Hauge}}, \bibinfo
  {author} {\bibfnamefont {S.}~\bibnamefont {Assali}}, \bibinfo {author}
  {\bibfnamefont {E.~P. A.~M.}\ \bibnamefont {Bakkers}}, \ and\ \bibinfo
  {author} {\bibfnamefont {M.}~\bibnamefont {Poggio}},\ }\href@noop {}
  {\bibfield  {journal} {\bibinfo  {journal} {Nature Phys.}\ }\textbf {\bibinfo
  {volume} {9}},\ \bibinfo {pages} {631} (\bibinfo {year} {2013})}\BibitemShut
  {NoStop}%
\bibitem [{\citenamefont {Slichter}(1980)}]{bib:Slichter}%
  \BibitemOpen
  \bibfield  {author} {\bibinfo {author} {\bibfnamefont {C.~P.}\ \bibnamefont
  {Slichter}},\ }\href@noop {} {\emph {\bibinfo {title} {Principles of Magnetic
  Resonance}}}\ (\bibinfo  {publisher} {Springer-Verlag, Berlin},\ \bibinfo
  {year} {1980})\BibitemShut {NoStop}%
\bibitem [{\citenamefont {White}(2007)}]{bib:White}%
  \BibitemOpen
  \bibfield  {author} {\bibinfo {author} {\bibfnamefont {R.}~\bibnamefont
  {White}},\ }\href@noop {} {\emph {\bibinfo {title} {Quantum Theory of
  Magnetism}}}\ (\bibinfo  {publisher} {Springer-Verlag, Berlin},\ \bibinfo
  {year} {2007})\BibitemShut {NoStop}%
\bibitem [{\citenamefont {Giamarchi}(2003)}]{giamarchi_book}%
  \BibitemOpen
  \bibfield  {author} {\bibinfo {author} {\bibfnamefont {T.}~\bibnamefont
  {Giamarchi}},\ }\href@noop {} {\emph {\bibinfo {title} {Quantum Physics in
  One Dimension}}}\ (\bibinfo  {publisher} {Oxford University Press, New
  York},\ \bibinfo {year} {2003})\BibitemShut {NoStop}%
\bibitem [{\citenamefont {Meng}\ \emph {et~al.}(2014)\citenamefont {Meng},
  \citenamefont {Stano}, \citenamefont {Klinovaja},\ and\ \citenamefont
  {Loss}}]{meng_13_nucl}%
  \BibitemOpen
  \bibfield  {author} {\bibinfo {author} {\bibfnamefont {T.}~\bibnamefont
  {Meng}}, \bibinfo {author} {\bibfnamefont {P.}~\bibnamefont {Stano}},
  \bibinfo {author} {\bibfnamefont {J.}~\bibnamefont {Klinovaja}}, \ and\
  \bibinfo {author} {\bibfnamefont {D.}~\bibnamefont {Loss}},\ }\href@noop {}
  {\bibfield  {journal} {\bibinfo  {journal} {Eur. Phys. J. B}\ }\textbf
  {\bibinfo {volume} {87}},\ \bibinfo {pages} {203} (\bibinfo {year}
  {2014})}\BibitemShut {NoStop}%
\bibitem [{\citenamefont {Abragam}(1961)}]{bib:Abragam}%
  \BibitemOpen
  \bibfield  {author} {\bibinfo {author} {\bibfnamefont {A.}~\bibnamefont
  {Abragam}},\ }\href@noop {} {\emph {\bibinfo {title} {The principles of
  nuclear magnetism}}}\ (\bibinfo  {publisher} {Clarendon Press, Oxford},\
  \bibinfo {year} {1961})\BibitemShut {NoStop}%
\bibitem [{\citenamefont {Maletinsky}\ \emph {et~al.}(2007)\citenamefont
  {Maletinsky}, \citenamefont {Badolato},\ and\ \citenamefont
  {Imamoglu}}]{Maletinsky}%
  \BibitemOpen
  \bibfield  {author} {\bibinfo {author} {\bibfnamefont {P.}~\bibnamefont
  {Maletinsky}}, \bibinfo {author} {\bibfnamefont {A.}~\bibnamefont
  {Badolato}}, \ and\ \bibinfo {author} {\bibfnamefont {A.}~\bibnamefont
  {Imamoglu}},\ }\href@noop {} {\bibfield  {journal} {\bibinfo  {journal}
  {Phys. Rev. Lett.}\ }\textbf {\bibinfo {volume} {99}},\ \bibinfo {pages}
  {056804} (\bibinfo {year} {2007})}\BibitemShut {NoStop}%
\bibitem [{\citenamefont {Sasaki}\ \emph {et~al.}(2014)\citenamefont {Sasaki},
  \citenamefont {Nonaka}, \citenamefont {Kunihashi}, \citenamefont {Kohda},
  \citenamefont {Bauernfeind}, \citenamefont {Dollinger}, \citenamefont
  {Richter},\ and\ \citenamefont {Nitta}}]{Sasaki}%
  \BibitemOpen
  \bibfield  {author} {\bibinfo {author} {\bibfnamefont {A.}~\bibnamefont
  {Sasaki}}, \bibinfo {author} {\bibfnamefont {S.}~\bibnamefont {Nonaka}},
  \bibinfo {author} {\bibfnamefont {Y.}~\bibnamefont {Kunihashi}}, \bibinfo
  {author} {\bibfnamefont {M.}~\bibnamefont {Kohda}}, \bibinfo {author}
  {\bibfnamefont {T.}~\bibnamefont {Bauernfeind}}, \bibinfo {author}
  {\bibfnamefont {T.}~\bibnamefont {Dollinger}}, \bibinfo {author}
  {\bibfnamefont {K.}~\bibnamefont {Richter}}, \ and\ \bibinfo {author}
  {\bibfnamefont {J.}~\bibnamefont {Nitta}},\ }\href@noop {} {\bibfield
  {journal} {\bibinfo  {journal} {Nature Nanotechnology}\ }\textbf {\bibinfo
  {volume} {9}},\ \bibinfo {pages} {703} (\bibinfo {year} {2014})}\BibitemShut
  {NoStop}%
\end{thebibliography}%
\end{document}